\documentclass[prl,twocolumn,superscriptaddress]{revtex4}
\usepackage{times}
\usepackage{graphicx}
\usepackage{amsmath}
\usepackage{hyperref}

\def\be{\begin{equation}}
\def\ee{\end{equation}}
\def\nn{\nonumber}
\def\psid{\psi^{\dagger}}
\begin{document}

\title{Non-Abelian Anyon Superconductivity}
\author{Waheb Bishara}
\affiliation{ Department of Physics, California Institute of
  Technology, MC 256-80 Pasadena, CA 91125}
\author{Chetan Nayak}
\affiliation{Microsoft Station Q, University of California,
Santa Barbara, CA 93106}
\affiliation{Department of Physics and Astronomy,
University of California, Los Angeles, CA 90095-1547}

\begin{abstract}
Non-Abelian Anyons exist in certain spin models and may exist in Quantuam Hall systems at certain filling fractions. In this work we studied the ground state of dynamical $SU(2)$ level-$\kappa$ Chern Simons non-Abelian anyons at finite density and no external magnetic field. We find that in the large-$\kappa$ limit the topological interaction induces a pairing instability and the ground state is a superconductor with  $\it{d+id}$ gap symmetry. We also develop a picture of pairing for the special value $\kappa=2$ and argue that the ground state is a superfluid of pairs for all values of $\kappa$.
\end{abstract}
\maketitle

\paragraph{Introduction.}
In two spatial dimensions, indistinguishable hardcore quantum particles are not restricted to being either fermions or bosons and can be \emph{anyons} \cite{stat}.
The simplest anyons have abelian statistics,
for which the wavefunction acquires a phase $\theta$ different from zero or $\pi$ when the positions of two particles are exchanged in a counter-clockwise manner,
thereby implementing a form of topological interaction.
Quasi-particles in many Quantum Hall states, such as $\nu=1/3$,
are expected to have such abelian fractional statistics \cite{Halperin84,Arovas84}.
A finite density of such fractional quasiparticles can form its own fractional quantum
Hall state, thereby leading to states such as $\nu=2/5$ (which also has
anyonic quasiparticle excitations above its ground state) \cite{Haldane83,Halperin84}.
Anyons with abelian fractional statistics also appear in microscopic models
of frustrated magnets in zero field \cite{abelian,Freedman04}, and one
interesting question to ask is what is the ground state of a
system with a finite density of anyons in zero field.
It was found that the ground state of abelian anyons which do not interact
with each other except through their topological interactions is a superfluid \cite{Fetter89,Chen89}. The physical picture is that the motion of a single anyon is frustrated since its phase gets scrambled due to the topological interactions with all the other anyons. Therefore, its kinetic energy is highly frustrated.
A collection of anyons might, for some values of $\theta$, behave like bosons and therefore may condense to form a superfluid. An argument based on
a mean-field treatment of the topological interaction supports this
conclusion \cite{Fetter89,Chen89}.

Recently, anyons with non-abelian statistics have been the subject of
great interest because of their potential application to topological
quantum computation \cite{tqc}. For these particles, there is a degeneracy of the ground state for given positions of the particles (so long as they are sufficiently
far apart) and adiabatically exchanging particles causes the
quantum state to rotate in this degenerate subspace.
Non-abelian anyons are hypothesized
\cite{Moore91,Greiter92,Nayak96c} to exist at the $\nu=\frac{5}{2}$ fractional quantum Hall state \cite{Willett87,Eisenstein02,Xia04}.
The vortices of a $p_x+ip_y$ superconductor can also have non-abelian statistics,
which may be observable in Sr$_2$RuO$_4$ \cite{Read00,Ivanov01,DasSarma06}.
What is the ground state of a finite density of non-abelian anyons
in zero magnetic field? The general argument about the frustration of kinetic
energy applies equally well to non-Abelian anyons. If a collection of non-Abelian anyons is bosonic, then it can condense, potentially lowering the kinetic
energy. However, the mean-field argument for Abelian anyons does
not apply to the non-Abelian case. Furthermore,
in order for this possibility to make sense,
the motion of the non-Abelian anyons must break the large
degeneracy of multi-quasiparticle states.
In this paper, we give arguments which show that,
at least in the case of non-abelian anyons described by $SU(2)_\kappa$
Chern-Simons gauge theory, a finite density of non-abelian anyons has a
superfluid ground state. In other words, we expect that doping a non-Abelian
topological phase in this class has the effect of driving a system
into a superconducting phase. We note that the case of non-Abelian anyons
coupled to both $SU(2)$ {\it and} $U(1)$ Chern-Simons fields has
also been shown to be superconducting when the gauge symmetry
is broken down to $U(1)$ \cite{Cappelli95}.

\paragraph{Non-Abelian Chern-Simons Anyons.}
The fractional statistics of anyons can be implemented in the boundary conditions of their wavefuntions, in which case the wavefunctions will be multivalued as a function of particle positions. The statistics can also be implemented by working with single-valued wavefunctions (i.e. bosonic or fermionic) with topological interactions between particles, and the statistics appear in the adiabatic transport of particles around each other
\cite{Verlinde91,LeeT94}.
In the abelian case, the topological interaction is introduced via a fictitious $U(1)$ gauge field with a Chern-Simons (CS) action interacting with the
particles \cite{stat}.
The gauge field effectively attaches fictitious magnetic flux to each particle, such that the wave function accumulates an Aharonov-Bohm phase whenever one particle winds around another.

Non-abelian fractional statistics can be similarly realized by particles interacting via a non-abelian gauge field with a CS action:
\be
\label{S_CS}
S_{CS}=\frac{\kappa}{4\pi} \int d^3x \,\epsilon^{\mu\nu\lambda}\, tr\left(A_\mu\partial_\nu A_\lambda+\frac{2}{3}A_\mu A_\nu A_\lambda\right)
\ee
where $A_\mu=A_\mu^a T^a$, $T^a$ are the generators of the gauge group, and $\kappa$ is an integer, referred to as the level of the theory. We limit our discussion to the gauge group $SU(2)$. With each particle interacting with this gauge field there is an associated representation of the group, its `isospin'.
This internal quantum number is totally distinct from the actual
spin of the particle. The representation carried by a particle or
collection of particles is physically meaningful, but a particular isospin
direction can be changed by a gauge transformation.
On a closed surface, the total isospin of all of the particles must be zero. However, even for given positions of particles in this theory, there is a non-trivial ground state degeneracy, corresponding to different ways in which the isospins can combine (or `fuse') to zero.
The allowed isospin states are a truncated version of the usual $SU(2)$ spin addition rules, and the truncation depends on the level $\kappa$. Adiabatically moving one particle around another rotates the quantum state of the system in this space of allowed fusion channels. Hence, the particles have non-abelian statistics.
Such non-abelian Chern-Simon anyons (NACS) appear as excitations of spin models in topological phases \cite{Freedman04,Levin05,Fendley05,Kitaev06}.
Therefore it is interesting to ask what is the ground state of a system
with a finite density of such particles that are not static and are able
to move around, a situation that can arise from doping spin systems
in topological phases.
For the sake of concreteness, we assume the particles are in the `isospin' $1/2$ representation.

\paragraph{Large-$\kappa$ Limit.}

The Hamiltonian of the CS gauge field vanishes, since the Lagrangian in (\ref{S_CS}) is linear in
time derivatives, but the gauge field itself does not vanish and is determined by the matter fields. In second quantized form the full Hamiltonian is \cite{Kim94,LeeT94}:
\be
H=\frac{1}{2m}\int d^2r \,D_i\psi^{\dagger} D_i\psi
\ee
where $D_i=\partial_i-iA_i^aT^a$, complemented with the constraints:
\be
\frac{1}{2}\epsilon^{ij}F^a_{ij}=\epsilon_{ij}\partial_i A_j^a+\frac{1}{2}\epsilon^{ij}f^{abc}A_i^b A_j^c = -\frac{1}{\kappa}\rho^a
\ee
$T^a,a=1,2,3$ are the generator of $SU(2)$, which are proportional to the Pauli matrices $\sigma^a$ in the spin $1/2$ representation. The matter field $\psi(r,t)$ is a fermionic spin-$1/2$ two-component spinor.

We consider large $\kappa$, assuming $A_i^a$ is analytic in $1/\kappa$.
To zeroth order, the $A$'s vanish; and to first order,
Gauss' law is linear and reduces to the abelian form:
\be
\epsilon_{ij}\partial_i A_j^a= -\frac{1}{k}\rho^a
\ee
This can be easily solved in the gauge $\nabla \cdot A^a=0$,
similar to the abelian case \cite{Greiter92}.
The Hamiltonian to first order in $1\over\kappa$ is:
\begin{align}
\label{firstorder}
H=& H_0+H_I \nn \\
H_0=& \frac{1}{2m}\int d^2r\,\partial_i\psi^{\dagger}\partial_i\psi \nn \\
H_I= &\frac{-1}{m}\int d^2r\, \psi^{\dagger}(\vec{A}^a \cdot i\vec{\nabla}) T^a \psi \nn \\
=&\frac{2\pi}{\kappa}\frac{i}{m}\sum_{k,k',q} \left( \frac{\vec{q}\times\vec{k}}{|\vec{q}|^2} T^a_{\beta\gamma}T^a_{\alpha\delta}\right) \psi^{\dagger}_{\alpha,k+q}\psi^{\dagger}_{\beta,k'-q}\psi_{\gamma,k'}\psi_{\delta,k}
\end{align}

The interaction is weak in the large $\kappa$ limit, and we may use the Renormalization Group (RG) scheme of Shankar \cite{Shankar94} to analyze it. The RG focuses on the fermionic modes that are near the Fermi surface and examines the flow of interaction strength in the different channels as the high energy modes are decimated. The one-loop flow reveals that only the BCS interaction is relevant if it is attractive in any of the angular momentum channels. We may write the BCS part of the interaction in (\ref{firstorder}) in terms of its angular momentum components (suppressing the isospin indices and the momentum sums) :
\begin{align*}
&\frac{2\pi}{\kappa}\frac{i}{m} \frac{\vec{k}\times \vec{k'}}{|\vec{k}-\vec{k'}|^2} \psid_k\psid_{-k}\psi_{-k'}\psi_{k'}  \nn \\
&\propto \frac{1}{m} \sum_{l\neq 0}  e^{il\theta} \frac{sgn(l)}{\kappa}\left(\lambda-\sqrt{\lambda^2-1})\right)^{|l|} \psid_k\psid_{-k}\psi_{-k'}\psi_{k'}
\end{align*}
where $\theta$ is the angle between $\vec{k}$ and $\vec{k'}$, and $\lambda=\frac{1}{2}\left(k/k' + k'/k\right) \geq 1$. Depending on the sign of $\kappa$, the interaction will be attractive in either the positive angular momentum channels or the negative one. Therefore it is relevant and there is a BCS pairing instability. This instability justifies using a BCS paired wavefunctions as a variational guess for the ground state wavefunction of the system, in which fermion bilinears $\langle \psid_{k,\alpha} \psid_{-k,\beta}\rangle$ develop a non zero expectation value. The variational parameters are the isospin structure and angular momentum dependence of these bilinear expectation values.

Following the usual BCS approach and assuming the order parameter has a definite angular momentum dependence, $\Delta_k=|\Delta_k|e^{i l \phi_k}$, and either a singlet or a triplet isospin structure, we obtain a self consistency gap equation for the order parameter:
\be
|\Delta_k|= \frac{C_{s/t}}{\kappa} \left[ \int_0^{k} \left( \frac{k'}{k}\right)^l \frac{|\Delta_{k'}|} {E_{k'}} k' dk'  +  \int_0^{k} \left( \frac{k}{k'}\right)^l \frac{|\Delta_{k'}|} {E_{k'}} k' dk'  \right]
\ee
The angular momentum $l$ must be even for spin singlet pairing and odd for spin triplet, and the interaction is effectively stronger in the singlet channel $C_s=3C_t$. We find that the lowest energy paired state is one with isospin singlet pairing and $l=2$ pair angular momentum. Since an isospin singlet does not couple to
the SU(2) gauge field, such a pair is a boson. If the anyons are charged, this is a superconductor with $\it{d+id}$ pairing.

One might worry that the BCS approach is not valid because the topological
interaction between quasiparticles is long-ranged. However, when two
$SU(2)_\kappa$ anyons pair to form a boson, the resulting bosons
do not have any remaining long-ranged interactions.
By pairing in singlets, the anyons lose their non-abelian interaction,
and the ground state can be thought of as condensation of pairs,
similar to condensation of pairs of fermions in superconductors.
Therefore, the long-ranged part of the topological interaction is not
important in this finite-density situation. At distances longer than the
superconducting coherence length, the topological interactions do not affect
the important degrees of freedom.

\paragraph{$\kappa=2$ Case.}
We now argue that a phase with excitations satisfying the $SU(2)_2$ NACS statistics can be reached by quantum disordering a superfluid, which implies the inverse:
$\kappa=2$ NACS anyons can condense into a superfluid.

To disorder a superfluid, it is useful to invoke a duality transformation and write the effective action in terms of vortex variables \cite{Fisher89a,Balents05,Wen90c}:
\begin{eqnarray}
\label{BosonicVortex}
S_v=\int d\tau d^2x \left(
\left|(\partial_\mu-ia_\mu)\varphi^v\right|^2  + V(|\varphi^v|^2)+ \right. \nonumber \\
\left. \frac{1}{g^2}(\epsilon_{\mu\nu\lambda}\partial_\nu a_\lambda)^2 + \frac{1}{2\pi}A^{em}_\mu \epsilon_{\mu\nu\lambda}\partial_\mu a_\lambda + ...
\right)
\end{eqnarray}
where $\varphi^v$ is the vortex annihilation operator. The vortices interact logarithmically, as expected, via the gauge field $a_\mu$, which is related to the super-current by:
$j_\mu=\frac{1}{2\pi}\epsilon_{\mu\nu\lambda}\partial_\nu a_\lambda$
and $A^{em}$ is the external electromagnetic potential that couples to the charge current. The superfluid and insulator phases can be recovered from this effective action: When the vortices are gapped, $\langle\varphi^v\rangle=0$, the vortex field and the gauge field $a_\mu$ can be integrated out and the resulting effective action contains a mass term for the external electromagnetic potential, i.e. the system is a superconductor. When vortices condense, $\langle\varphi^v\rangle\neq 0$, the resulting effective action for $A^{em}$ has a Maxwell term to leading order, so the system is insulating. Condensation of any powers of the vortex field would also result in an insulator.

Recently, it has been argued that it is justified in certain limits to treat vortices as weakly interacting fermions rather than strongly interacting bosons \cite{Alicea05,Galitski05}. Fermionizing is achieved by attaching one Abelian
Chern-Simons flux quantum per particle. This procedure makes vortices fermionic at the expense of adding a gauge field $\alpha_\mu$ with a CS action:
\begin{align}
\label{fermionicvortices}
{\cal L}_v&= \bar{\psi}\gamma^\mu(\partial_u-ia_\mu-i\alpha_\mu)\psi + V(\bar{\psi}\psi) \nonumber \\
& + \frac{1}{2g^2} (\epsilon_{\mu\nu\lambda}\partial_\nu a_\lambda)^2+ \frac{i}{4\pi}\epsilon_{\mu\nu\lambda}\alpha_\mu\partial_\nu\alpha_\lambda + \frac{1}{2\pi}A^{em}_\mu j_\mu
\end{align}
where $\psi$ is the usual two-component spinor of Dirac fermions in $2+1$ dimensions
and the $\gamma^\mu$ can be chosen to be Pauli matrices. In the presence of a gauge field with a Maxwell action, it is possible to integrate out the CS field and this will generate only terms that are higher order than the Maxwell term for the remaining gauge field; naively, they are less relevant. This is done by shifting
$a_\mu\rightarrow{a_\mu}-{\alpha_\mu}$ to decouple ${\alpha_\mu}$ from
the fermions. Then ${\alpha_\mu}$ can be integrated out to obtain:
\begin{align}
\label{shiftedL_v}
{\cal L}_v&= \bar{\psi}\gamma^\mu(\partial_u-ia_\mu)\psi + m(\bar{\psi}\psi) + G(\bar{\psi}\psi)^2 \nonumber \\
& + \frac{1}{2g^2} (\epsilon_{\mu\nu\lambda}\partial_\nu a_\lambda)^2 + \frac{1}{2\pi}A^{em}_\mu j_\mu + O(\partial^3 a^2)
\end{align}
where we have expanded $V(\bar{\psi}\psi)$. We will be interested in a phase in which
spontaneous symmetry breaking leads to a mass for the gauge field $a_\mu$,
and therefore higher-order terms in ${a_\mu}$ can be treated perturbatively.
We ignore them here.


In terms of fermionic vortices, an insulating phase is reached when pairs of vortices condense. Pairing of fermions with a relativistic spectrum has been explored in two \cite{Ohsaku04} and three \cite{Capelle97} spatial dimensions. We can decouple the quartic interaction term in (\ref{shiftedL_v}) via a Hubbard-Stratonovich transformation and look for a self consistent pairing $\mathbf{\Delta}_{ij}$:
\begin{align}
\label{pairedaction}
S_{\rm f}=\int d\tau &d^2x  \, \bar{\psi}\gamma^\mu(\partial_u-ia_\mu)\psi + m\bar{\psi}\psi + \nonumber \\& +\psi^{\dagger}(x)_i\mathbf{\Delta}_{ij}(x)\psi^{\dagger}(x) + h.c.
\end{align}
It was found \cite{Ohsaku04} that for $G>0$ no such pairing exists, while for $G<0$ the only self consistent pairing is $\mathbf{\Delta }= \gamma^2 \Delta(x)$, where $\Delta(x)$ is a scalar. When such pairing happens, the system is in the electrically insulating phase.


In the presence of pairing, the mean field Hamiltonian is quadratic and is diagonalized by fermionic operators, $a_{jk}$, that are linear combinations of $\psi_i$ and $\psi^{\dagger}_j$ \cite{Capelle97}:

\be
\label{lineartrans}
\psi_i(x)=\sum_{i,j=1,2} \left( u_{ijk}(x)a_{jk} + v^*_{ijk}(x)a^{\dagger}_{jk}\right)
\ee
the $i$ and $j$ indices are component indices, and the index $k$ enumerates the state, for example it can be momentum in a homogeneous system. The functions $u_{ijk}(x)$ and $v_{ijk}(x)$ solve the Bogoliubov deGennes (BdG) equations:

\be
\label{RBdG}
\left(
\begin{array}{cc}
\hat{h_D} & \gamma^2 \Delta(x)\\
-(\gamma^2 \Delta(x))^* & -\hat{h_D}^*
\end{array}
\right)
\left(
\begin{array}{c}
u_1(x) \\ u_2(x)\\v_1(x)\\v_2(x)
\end{array}
\right)
= E_{jk} \left(
\begin{array}{c}
u_1(x) \\ u_2(x)\\v_1(x)\\v_2(x)
\end{array}
\right)
\ee
with $\hat{h_D}=\gamma^0(-i\vec{\nabla}\cdot\vec{\gamma} + m)$ the Dirac Hamiltonian. Equations of similar structure have been discussed in ref. \cite{HouChamonMudry} in the context of graphene, with the important difference that the order parameter there corresponds to a mass rather than pairing of the fermions.

The pairing $\Delta(x)$ is much like the familiar pairing of non-relativistic fermions. Most importantly, it gives the dual gauge field a mass, and the system is a ``dual superconductor'', i.e.
an electric insulator. The effective action for $\Delta(x)$ is simply a Landau-Ginzburg free energy functional of a three dimensional superconductor, and we will draw on our knowledge of the conventional theory of superconductivity to learn about the phase described above.

The boson charge density maps under duality to the
dual magnetic flux $j_0 = \frac{1}{2\pi}\vec{\nabla}\times\vec{a}$.
so changing the charge density is equivalent to introducing dual magnetic flux to the system. In the insulating phase where there is vortex pairing, the dual magnetic flux can form a lattice of defects in the vortex pair order parameter $\Delta(x)$, each carrying half a unit of dual flux, i.e. an Abrikosov lattice. This configuration corresponds to a crystal of localized charges, each carrying half the charge of the original bosons which formed the superconductor. If the bosons are charge $2e$ Cooper pairs, then the paired-vortex insulating phase has localized excitations of charge $e$. We will now show that these excitations also obey $SU(2)_2$ NACS statistics.

A prerequisite for having non-abelian statistics is having a degeneracy for a given position of the excitations. In the paired-vortex insulator described above, the degeneracy is due to zero modes localized inside the defects in the pairing field $\Delta(x)$. For simplicity, a defect can be modeled as a region where the magnitude of the pairing field $\Delta(x)$ drops abruptly to zero and its phase changes by $2\pi$ in going around the defect,
$\Delta(r,\theta)=\Theta(r-r_0)\Delta_\infty e^{\pm i\theta}$
with $\Theta(s)$ the step function, and $r_0$ is on the order of the coherence length of the condensate, $\xi_v$. Solving the BdG equations, eq.(\ref{RBdG}), with this $\Delta(r,\theta)$
gives the spectrum of excitations in the presence of a defect. We are interested in solutions at zero energy, and if we also restrict ourselves to solutions satisfying $u_i(x)=v^*_i(x)$, then the BdG equations reduce to only two linearly independent equations:
\begin{align}
mu_1+e^{-i\theta}(\partial_r-\frac{i}{r}\partial_\theta)u_2+\Delta(r)e^{\pm i \theta}u_2^* & =0  \nn \\
-e^{i\theta}(\partial_r+\frac{i}{r}\partial_\theta)u_1-mu_2-\Delta(r)e^{\pm i\theta}u_1^* & =0
\end{align}
we can assume $m>0$ and $\Delta_\infty>0$ without loss of generality. These equations can be solved exactly, but it is enough to examine their behavior for large $r$. The asymptotic behavior of $u_1$ and $u_2$ is:
\be
\label{asymptotic}
u_{1,2}(r \rightarrow \infty)  \sim Ae^{-(\Delta_\infty+m)r}+Be^{-(\Delta_\infty-m)r}
\ee
For $\Delta_\infty <m$, there is no well behaved localized solution, but for $\Delta_\infty >m$ there is one solution which decays exponentially away from the defect.
The fermionic operator $a^{\dagger}\equiv \gamma^{\dagger}$ corresponding to this
zero-energy solution localized at the defect has the property $\gamma^{\dagger}=\gamma$,
i.e. it is a Majorana fermion, as can be seen from eq.(\ref{lineartrans}) with $u_i(x)=v^*_i(x)$.
A two-state Hilbert space cannot be built from a single Majorana fermion, but in the presense of two defects in $\Delta(x)$, which correspond to two localized charges that are well separated, there is a Majorana mode in each of the defects, which allows us to define a Dirac fermion $\Psi=\frac{1}{\sqrt{2}}(\gamma_1+i\gamma_2)$
satisfying the usual anticommutation rules, $\{\Psi,\Psi^{\dagger}\}=1$.
The occupation number of the fermionic mode $\Psi$ defines two degenerate zero energy states, $|0\rangle$ and $\Psi^{\dagger}|0\rangle$. This degeneracy does not correspond to a local quantum number since the Dirac fermion is shared between the defects, regardless of their separation. For $2N$ defects in the the pairing field $\Delta(x)$ there are $2^N$ degenerate quantum states at zero energy. The non-abelian statistics is evident when exchanging positions of defects. When defects $i$ and $j$ are exchanged, the Majorana zero modes are transformed by $\gamma_i\rightarrow \gamma_j$ and $\gamma_j\rightarrow -\gamma_i$, which can be translated into a rotation in the $2^N$ dimensional degenerate subspace of states \cite{Ivanov01,Stern04}. This type of non-abelian statistics corresponds to $SU(2)_2$.

\vspace{2mm}
\paragraph{Discussion.}
In this paper, we have studied the ground state of a finite density of dynamical
non-abelian SU(2)$_\kappa$ Chern-Simons anyons. In the limit $\kappa \rightarrow \infty$ we showed that the topological interaction between anyons induces a pairing instability, and the preferred pairing has a $\it{d+id}$ spatial symmetry and $j=0$ isospin structure. For the special value $\kappa=2$, we showed that a superfluid of pairs can be disordered to get an
insulator with SU(2)$_2$ anyonic excitations, which implies the inverse path is also possible,
and SU(2)$_2$ anyons can pair to form a superfluid. Since pairs of NACS anyons in the $j=0$ channel are bosonic with respect to each other {\it for all values of $\kappa$}, we suggest that the ground state for a finite density of NACS anyons is a superfluid of bosonic pairs
for all $\kappa$.

The most obvious implication is that if a frustrated magnet were found in
a non-Abelian topological phase, then doping the system would drive it superconducting.
Conversely, non-Abelian anyon superconductivity may be an explanation
for the superconductivity of some anomalous materials. For the route to superconductivity
which we sketched above for $\kappa=2$, vortex pairs condense en route to
the non-Abelian insulating state. Consequently, individual vortices do not
condense. Therefore, flux-trapping experiments \cite{Senthil01a} should
find a positive result for such a superconductor.

We would like to thank P. Bonderson, E. Demler and K. Shtengel for helpful comments and discussions. This work has been supported by the NSF under grant
DMR-0411800 and the ARO under grant W911NF-04-1-0236.


\begin{thebibliography}{99}

\bibitem{stat} J. Leinaas and J. Myrheim, Nuovo Cimento {\bf 37B}
  (1977) 1;  F. Wilczek, Phys. Rev. Lett. {\bf 48} (1982);
  {\it ibid}. {\bf 49} (1982) 957. See also F. Wilczek, ed. {\it Fractional Statistics and
    Anyon Superconductivity} (World Scientific, Singapore 1990).


\bibitem{Halperin84}
B.~I. Halperin, Phys. Rev. Lett. {\bf 52},  1583  (1984).

\bibitem{Arovas84}
D. Arovas, J.~R. Schrieffer, and F. Wilczek, Phys. Rev. Lett. {\bf 53},  722
  (1984).

\bibitem{Haldane83}
F.~D.~M. Haldane, Phys. Rev. Lett. {\bf 51},  605  (1983).

\bibitem{abelian}
V. Kalmeyer and R.~B. Laughlin, \prl {\bf 59}, 2095 (1987).

\bibitem{Freedman04}
M. Freedman {\it et al.}, Ann. Phys. {\bf 310}, 428 (2004).

\bibitem{Fetter89}
A. Fetter, C. Hanna, and R. Laughlin, Phys. Rev. B {\bf 39},  9679  (1989).

\bibitem{Chen89}
Y. Chen, F. Wilczek, E. Witten, and B. Halperin, Int. J. Mod. Phys. B {\bf 3},
  1001  (1989).

\bibitem{tqc} S. Das Sarma, M. Freedman, and C. Nayak,
{\it Physics Today} {\bf 59}, 32 (2006) and references therein.

\bibitem{Moore91}
G. Moore and N. Read, Nucl. Phys. B {\bf 360},  362  (1991).

\bibitem{Greiter92}
M. Greiter, X.~G. Wen, and F. Wilczek, Nucl. Phys. B {\bf 374},  567  (1992).

\bibitem{Nayak96c}
C. Nayak and F. Wilczek, Nucl. Phys. B {\bf 479},  529  (1996).

\bibitem{Willett87}
R. Willett {\it et~al.}, Phys. Rev. Lett. {\bf 59},  1776  (1987).

\bibitem{Xia04}
J.~S. Xia {\it et~al.}, Phys. Rev. Lett. {\bf 93},  176809  (2004).

\bibitem{Eisenstein02}
J.~P. Eisenstein, K.~B. Cooper, L.~N. Pfeiffer, and K.~W. West, Physical Review
  Letters {\bf 88},  076801/1  (2002).

\bibitem{Read00}
N. Read and D. Green, \prb {\bf 61}, 10267 (2000).

\bibitem{Ivanov01}
D.~A. Ivanov, Phys. Rev. Lett. {\bf 86},  268  (2001).

\bibitem{DasSarma06}
S. Das~Sarma, C. Nayak, and S. Tewari, Phys. Rev. B {\bf 73},  220502
  (2006).

\bibitem{Cappelli95}
  A. Cappelli and P. Valtancoli, Nucl. Phys. B {\bf 453}, 727 (1995).

\bibitem{LeeT94}
T. Lee and P. Oh, Phys. Rev. Lett. {\bf 72},  1141  (1994).

\bibitem{Levin05}
M.~A. Levin and X.-G. Wen, Phys. Rev. B {\bf 71}, 045110 (2005).

\bibitem{Fendley05}
P. Fendley and E. Fradkin, Phys.Rev. B{\bf 72}, 024412 (2005).

\bibitem{Kitaev06}
A. Kitaev, Ann. Phys. {\bf 321}, 2 (2006).

\bibitem{Shankar94}
R. Shankar, Rev. Mod. Phys. {\bf 66},  129  (1994).

\bibitem{Fisher89a}
M.~P.~A. Fisher and D.~H. Lee, Phys. Rev. B {\bf 39},  2756  (1989).

\bibitem{Wen90c}
X.~G. Wen and A. Zee, Int. J. Mod. Phys. B {\bf 4},  437  (1990).

\bibitem{Verlinde91}
E. Verlinde, in {\it Modern Quantum FIeld Theory},
(World Scientific, Singapore, 1991).

\bibitem{Kim94}
W.~T. Kim and C. Lee, \prd {\bf 49}, 6829 (1994).

\bibitem{Ohsaku04}
T. Ohsaku, Intl. J. Mod. Phys, B{\bf 18}, 1771 (2004).

\bibitem{Capelle97}
K. Capelle and E.~K.~U. Gross, \prb {\bf 59}, 7140 (1999).

\bibitem{Balents05}
L. Balents {\it et al.}, cond-mat/0504692.

\bibitem{Alicea05} J. Alicea, O. I. Motrunich, M. Hermele, and M. P. A. Fisher
Phys. Rev. B 72, 064407 (2005).

\bibitem{Galitski05}
V.~M. Galitski, G. Refael, M.~P.~A. Fisher, and T. Senthil, Phys. Rev. Lett.
  {\bf 95},  077002  (2005).

\bibitem{HouChamonMudry}
C.~Y. Hou, C. Chamon, and C. Mudry, cond-mat/0609740.

\bibitem{Stern04}
A. Stern, F. von Oppen, and E. Mariani, Phys. Rev. B {\bf 70},  205338  (2004).

\bibitem{Senthil01a}
T. Senthil and M.~P.~A. Fisher, Phys. Rev. Lett. {\bf 86},  292  (2001).

\end{thebibliography}

\end{document}